\let\ifarxiv=\iffalse    
\pdfoutput=1

\ifarxiv

\documentclass[12pt,a4paper]{article}
\usepackage[a4paper,text={450pt,650pt},centering]{geometry}

\fi

\ifarxiv\else

\documentclass[11pt,a4paper]{article}
\usepackage{mathptmx}
\usepackage[a4paper,text={130mm,198mm}]{geometry}

\fi

\usepackage{amsmath,amssymb}
\usepackage[bookmarks=true,hyperfigures=true]{hyperref}
\usepackage{graphicx}
\usepackage[nosort]{cite}
\ifarxiv\usepackage[bulletsep]{collref}\fi

\usepackage{amsfonts}
\usepackage{amscd}
\usepackage{bbm}

\let\oldbfseries=\bfseries
\let\oldmdseries=\mdseries
\let\oldnormalfont=\normalfont
\renewcommand{\bfseries}{\oldbfseries\boldmath}
\renewcommand{\mdseries}{\oldmdseries\unboldmath}
\renewcommand{\normalfont}{\oldnormalfont\unboldmath}

\allowdisplaybreaks[3]

\numberwithin{equation}{section}

\usepackage[font=small,labelfont=bf,width=0.85\textwidth]{caption}

\providecommand{\hypersetup}[1]{}
\providecommand{\texorpdfstring}[2]{#1}

\hypersetup{plainpages=false}
\hypersetup{pdfpagemode=UseNone}
\hypersetup{bookmarksnumbered=true}
\hypersetup{pdfstartview=FitH}
\hypersetup{colorlinks=false}
\hypersetup{citebordercolor={.5 1 .5}}
\hypersetup{urlbordercolor={.5 1 1}}
\hypersetup{linkbordercolor={1 .7 .7}}

\providecommand{\href}[2]{#2}
\providecommand{\arxivlink}[1]{\href{http://arxiv.org/abs/#1}{arxiv:#1}}

\def\[{\begin{equation}}
\def\]{\end{equation}}
\def\<{\begin{eqnarray}}
\def\>{\end{eqnarray}}

\def\({\left(}
\def\){\right)}

\def\AdSxS{{AdS${}_5 \times {}$S${}^5$}}

\newcommand{\comm}[2]{[#1,#2]}
\newcommand{\gen}[1]{\mathfrak{#1}}

\newcommand{\Ya}[1]{\mathcal{Y}(#1)}

\newcommand{\half}{\frac{1}{2}}

\newcommand{\copro}{\Delta}

\newcommand{\genY}[1]{\widehat{\mathfrak{#1}}}

\newcommand{\struc}{f}
\newcommand{\earel}[1]{\mathrel{}&\hspace{-2\arraycolsep}#1\hspace{-2\arraycolsep}&\mathrel{}}
\newcommand{\eq}{\earel{=}}
\newcommand{\alg}[1]{\mathfrak{#1}}

\newcommand{\mathsym}[1]{{}}

\renewcommand{\[}{\left[}
\renewcommand{\]}{\right]}

\begin{document}


\thispagestyle{empty}
\phantomsection
\addcontentsline{toc}{section}{Title}

\begin{flushright}\footnotesize%
\texttt{\arxivlink{1012.4005}}\\
overview article: \texttt{\arxivlink{1012.3982}}%
\vspace{1em}%
\end{flushright}

\begingroup\parindent0pt
\begingroup\bfseries\ifarxiv\Large\else\LARGE\fi
\hypersetup{pdftitle={Review of AdS/CFT Integrability, Chapter VI.2: Yangian Algebra}}%
Review of AdS/CFT Integrability, Chapter VI.2:\\
Yangian Algebra
\par\endgroup
\vspace{1.5em}
\begingroup\ifarxiv\scshape\else\large\fi%
\hypersetup{pdfauthor={Alessandro Torrielli}}%
Alessandro Torrielli
\par\endgroup
\vspace{1em}
\begingroup\itshape
Institute for Theoretical Physics and Spinoza Institute,\\
Utrecht University, 3508 TD Utrecht, The Netherlands
\par\vspace{1ex}
\quad and
\par\vspace{1ex}
Department of Mathematics, University of York\\
Heslington, York, YO10 5DD, U.K.
\par\endgroup
\vspace{1em}
\begingroup\ttfamily
alessandro.torrielli@york.ac.uk
\par\endgroup
\vspace{1.0em}
\endgroup

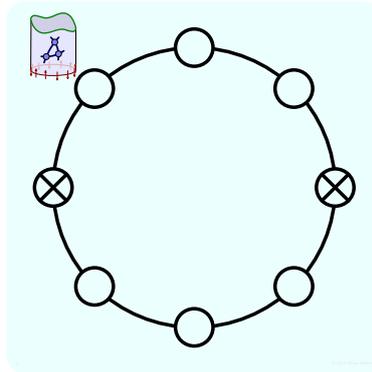
\begin{center}
\includegraphics[width=5cm]{TitleVI2.mps}
\vspace{1.0em}
\end{center}

\paragraph{Abstract:}
We review the study of Hopf algebras, classical and quantum R-matrices, infinite-dimensional Yangian symmetries and their representations in the context of integrability for the ${\cal{N}}=4$ {\it vs} $AdS_5 \times S^5$ correspondence.

\ifarxiv\else
\fi
\hypersetup{pdfsubject={MSC (2010): 81T60, 81T30, 81U15, 16T05, 17B65}}%

\ifarxiv\else
\fi
\hypersetup{pdfkeywords={ADS/CFT, Supersymmetry, Integrability, Exact S-matrix, Hopf algebras, Quantum groups, Yangians}}%

\newpage


\section{Introduction}

Despite the success obtained so far by the integrability program,
many questions are left unanswered. Most notably, the problem
remains of understanding what is the non-perturbative definition
of the model that seems to reproduce so well all the available
data \cite{Stau}. Answering this question may also be important
for a deeper understanding of the finite-size problem and its
solution. An essential role in this respect is played by the
symmetries of the factorized S-matrix. A clear sign is the
presence of a Hopf algebra
\cite{Janik:2006dc,Gomez:2006va,Plefka:2006ze}, then promoted to a
Yangian \cite{Beisert:2007ds}. In relativistic integrable quantum
field theories, symmetries like the Yangian or quantum affine
algebras completely determine the tensorial part of the S-matrix,
up to an overall scalar factor. They also entail important
consequences for the transfer matrices and for the Bethe equations
\cite{KR}. This happens also in the AdS/CFT case
\cite{deLeeuw:2008ye,Arutyunov:2009mi}. However, the AdS/CFT
Yangian has very distinctive features still preventing a full
mathematical understanding. For instance, there exists an
additional Yangian symmetry of the S-matrix
\cite{Matsumoto:2007rh,Beisert:2007ty} with properties not yet
entirely understood, pointing to a new type of quantum
group\footnote{The relation with Yangian symmetry in $n$-p.t
amplitudes \cite{Drummond:2010km} is also a fascinating problem.}. In
order to give an ultimate solution of the AdS/CFT integrable
system, one needs to understand the features of this novel quantum
group, and of the associated quantum integrable model. The scope
of this review is illustrating such group-theory aspects.

\section{Hopf Algebras}\label{Hop}
Let us begin by recalling a few concepts in the theory of Hopf algebras, as these are very important algebraic structures appearing in the context of integrable models. We will attempt to motivate these concepts mostly from the physical viewpoint, and refer the reader to standard textbooks, such as \cite{Chari}, for a thorough treatment.

The starting point is the algebra of symmetries of a system. Let us consider the case when this algebra is a Lie (super)algebra $\alg{g}$, and let us also consider its universal enveloping algebra $A\equiv U(\alg{g})$. This step allows us to `multiply' generators, besides taking the Lie bracket. In such universal enveloping algebra there is a {\it unit element} $\mathbbmss{1}$ with respect to the {\it multiplication map} $\mu$. We think about multiplication as $\mu: A \otimes A \rightarrow A$, and we introduce a {\it unit map} $\eta: \mathbbmss{C} \rightarrow A$. A few compatibility conditions on these maps guarantee that we are dealing with the physical symmetries of, say, a single-particle system.  

In order to treat multiparticle states, we equip our algebra with two more maps, and obtain a {\it bialgebra} structure. One map is the {\it coproduct} $\Delta: A \rightarrow A \otimes A$, which tells us how symmetry generators act on two-particle states. The other map is the {\it counit} $\epsilon: A \rightarrow \mathbbmss{C}$. A list of compatibility axioms ensures that these maps are consistent with the (Lie) (super)algebra structure, so we can safely think of them as the symmetries we started with, just acting on a Fock space. In fact, for a generic $n$-particle state, we can generalize the action of the coproduct as the composition $\Delta^n = ...(\Delta \otimes \mathbbmss{1} \otimes \mathbbmss{1})(\Delta \otimes \mathbbmss{1})\Delta$. The {\it coassociativity} axiom 
\begin{equation}
(\Delta \otimes \mathbbmss{1})\Delta \, = \,  (\mathbbmss{1} \otimes \Delta)\Delta 
\end{equation}
guarantees that a change in the positions of the $\Delta$'s in the sequence $\Delta^n$ is immaterial. 

One more map turns our structure into a {\it Hopf algebra}. This map is the {\it antipode} $\Sigma: A \rightarrow A$, which is needed to define antiparticles (conjugated representations of the symmetry algebra). Therefore, the antipode should also be consistent with the (Lie) (super)algebra structure\footnote{Being the antipode connected to conjugation, one imposes $\Sigma (ab) = (-)^{ab} \Sigma(b)\Sigma(a)$, where multiplication is {\it via} the map $\mu$.}, and be compatible with the coproduct action. If a bialgebra admits an antipode, it is unique.

In the scattering theory of integrable models, the fundamental object encoding the dynamics is the two-particle S-matrix, which exchanges the momenta of the two particles, and reshuffles their colors. One has therefore the possibility of defining the coproduct action as acting on, say, {\it in} states. Likewise, the composed map $P \Delta \equiv \Delta^{op}$, with $P$ the permutation map, will act on {\it out} states. The discovery of quantum groups revealed that these two actions need not be the same. They are the same only for {\it cocommutative} Hopf algebras, one example being the Leibniz rule $\Delta (a) = a \otimes \mathbbmss{1} + \mathbbmss{1} \otimes a$ one normally associates with local actions. In general, coproducts can be more complicated, as we will amply see in what follows\footnote{This is another reason why the Coleman-Mandula theorem does not apply to the S-matrices we will be discussing (besides being in $1+1$ dimensions).}.

However, as $\Delta$ and $\Delta^{op}$ produce tensor product representations of the same dimensions, they may be related by conjugation {\it via} an invertible element (the S-matrix itself). The Hopf-algebra is then said to be {\it quasi-cocommutative}, and, if the S-matrix satisfies an additional condition  (`bootstrap' \cite{Zamolodchikov:1978xm,Dorey:1996gd}), it is called {\it quasi-triangular}. The S-matrix must also be compatible with the antipode map, a condition that in physical terms goes under the name of {\it crossing} symmetry. One can prove that bootstrap implies that the S-matrix satisfies the Yang-Baxter equation and the crossing condition. 

As one can easily realize, the framework of Hopf algebras is particularly suitable for dealing with integrable scattering. Integrability reduces the scattering problem to an algebraic procedure, and the axioms we have been discussing just formalize that procedure. However, instead of being a mere translation, the mathematical framework of Hopf algebras provides a set of powerful theorems that unify the treatment of arbitrary representations. To this purpose, the notion of {\it universal R-matrix} is very important. This is an abstract solution to the quasi-cocommutativity condition, purely expressed in terms of algebra generators. This solution gives an expression for the S-matrix which is therefore free from a particular representation, at the same time being valid in any of them upon plug-in. As we will explicitly see in what follows, the study of the properties of the universal R-matrix reveals a big deal about the structure of the (hidden) symmetry algebra of the integrable system.

\section{Yangians}\label{ssec:drinf1}
Let $\alg{g}$ be a finite dimensional simple Lie algebra with generators
$\gen{J}^A$, structure constants $f^{AB}_C$ defined by $\comm{\gen{J}^A}{\gen{J}^B} = f^{AB}_C\gen{J}^C$ and a non-degenerate invariant bilinear form $\kappa^{AB}$. The Yangian $\Ya{\alg{g}}$ of $\alg{g}$
is a deformation of the universal enveloping algebra
of half of the loop algebra of $\alg{g}$. The loop algebra is defined by (\ref{loopa}), "half" meaning non-negative indices $m,n$.
Drinfeld gave two isomorphic realizations of the Yangian\footnote{The reader is referred to {\it e.g.}
\cite{Chari,Etingof,MacKay:2004tc,Molev} for a thorough treatment. We will not discuss the `RTT' realization, see {\it e.g} \cite{Faddeev:1987ph,Molev}. For generalizations to Lie
superalgebras, see {\it e.g.} \cite{Curtright:1992kp,Yao,stuko,Gow,Spill:2008yr,Rej:2010mu}.}.
The first realization \cite{Drin} is as follows.
$\Ya{g}$ is defined \nopagebreak by
relations between level zero generators $\gen{J}^A$ and level one generators $\genY{J}^A$: \<
\comm{\gen{J}^A}{\gen{J}^B} = f^{AB}_C\gen{J}^C,\qquad \qquad
\label{rels}\comm{\gen{J}^A}{\genY{J}^B} = f^{AB}_C\genY{J}^C. \> The
generators of higher levels are derived recursively by
computing the commutant, subject to the following Serre
relations (for $\alg{g} \neq \alg{su(2)}$): \<
\label{Serr}\comm{\genY{J}^{A}}{\comm{\genY{J}^B}{\gen{J}^{C}}} +
\comm{\genY{J}^{B}}{\comm{\genY{J}^C}{\gen{J}^{A}}} +
\comm{\genY{J}^{C}}{\comm{\genY{J}^A}{\gen{J}^{B}}} \eq
\frac{1}{4} f^{AG}_{D}f^{BH}_Ef^{CK}_{F}f_{GHK}
\gen{J}^{\{D}\gen{J}^E\gen{J}^{F\}}. \> Indices are raised (lowered)
with $\kappa^{AB}$ (its inverse). The Yangian is
equipped with a Hopf algebra structure. The coproduct is uniquely
determined for all generators by specifying it on the level zero
and one generators as follows:
\begin{equation}
\label{cop}
\copro (\gen{J}^A) =\gen{J}^A\otimes \mathbbmss{1}+\mathbbmss{1}\otimes\gen{J}^A, \qquad \copro( \, \genY{J}^A )=\genY{J}^A\otimes \mathbbmss{1}+\mathbbmss{1}\otimes\genY{J}^A+\half
\struc^{A}_{BC}\gen{J}^B\otimes\gen{J}^C.
\end{equation}
Antipode and counit are easily obtained
from the Hopf algebra definitions\footnote{{\it Via} a rescaling of the algebra generators, one can make a parameter (say, $\hbar$) appear in front of the mixed term $\half
\struc^{A}_{BC}\gen{J}^B\otimes\gen{J}^C$ in the Yangian coproduct (\ref{cop}). This parameter is sometimes useful as it can be made small, as in the classical limit, cf. section 3.}. We will not present here Drinfeld's second realization of the Yangian \cite{Dsecond}, which is suitable for constructing the universal
R-matrix \cite{Khoroshkin:1994uk}. It suffices to say that it explicitly solves the recursion implicit in the first realization.

\subsection{The \texorpdfstring{$\alg{psu}(2,2|4)$}{psu(2,2|4)} Yangian}

Generically, the level zero local generators are realized on
spin-chains as
\begin{eqnarray}
\gen{J}^A \, = \, \sum_k \, \gen{J}^A (k), \qquad k \in \{spin-chain \, \, \, \, \, sites\}.
\end{eqnarray}
For infinite length, the level one Yangian generators are bilocal combinations
\begin{eqnarray}
\label{lev1}
\genY{J}^A \, = \, \sum_{k<n} \, \struc^{A}_{BC} \, \gen{J}^B (k) \, \gen{J}^C (n).
\end{eqnarray}
The relationship with the coproduct (\ref{cop}) will be clear later when discussing the Principal Chiral Model.
Level $n$ generators are $n+1$-local expressions. At finite length, boundary effects usually prevent from
having conserved charges such as (\ref{lev1}), while Casimirs of the Yangian may still be well-defined.
We refer to \cite{Bernard:1992ya} for a review.

The ${\cal{N}}=4$ SYM spin-chain is based on the superconformal
symmetry algebra $\alg{psu}(2,2|4)$. The Yangian charges for
infinite length have been constructed, at leading order in the 't
Hooft coupling, in \cite{Dolan:2003uh}. The Serre relations for
the relevant representations have been proved in
\cite{Dolan:2004ps}. In \cite{Dolan:2004ys} the first two Casimirs
of the Yangian are computed and identified with the first two
local abelian Hamiltonians of the spin-chain with periodic
boundary conditions.

Perturbative corrections to the Yangian charges in subsectors have
been studied in
\cite{Serban:2004jf,Agarwal:2004sz,Agarwal:2005ed,Zwiebel:2006cb,Beisert:2007sk}.
The integrable structure of spin-chains with long-range (LR)
interactions, like the one emerging from gauge perturbation
theory, lies outside the established picture
\cite{Rej:2010ju}, but a large class of LR spin-chains has been shown
to display Yangian symmetries, see also
\cite{Beisert:2003tq,Beisert:2005wv,Beisert:2007jv,Zwiebel:2008gr}.
In absence of other standard tools, Yangian symmetry provides a
formal proof of integrability order by order in perturbation
theory. The two-loop expression of the
Yangian (\ref{lev1}) for the $\alg{su}(2|1)$ sector has
been derived in \cite{Zwiebel:2006cb}. In \cite{Beisert:2007sk}, a
large degeneracy of states in the $\alg{psu}(1,1|2)$ sector is
explained {\it via} nonlocal charges related to the loop-algebra
of the $\alg{su}(2)$ automorphism of $\alg{psu}(1,1|2)$. Further
references include
\cite{Agarwal:2004cb,Agarwal:2006nv}. For a recent review we recommend \cite{Beisert:2010jq}.

Higher non-local charges analogous to (\ref{lev1}) emerge in 2D
classically integrable field theories \cite{Magro:2010jx,SchaferNameki:2010jy}. If not anomalous, their quantum versions
\cite{Luscher:1977uq} form a
Yangian.
{\it E.g.}, for
the Principal Chiral
Model
\begin{eqnarray}
\label{nnl}
\frac{d}{dt} \, \widehat{\alg{J}}^A =  \frac{d}{dt} \int_{-\infty}^{\infty} dx \, \Big[ \epsilon_{\mu \nu} \, J^{\nu, \, A} + \frac{1}{2} \, f^A_{BC} \, J_\mu^B \int_{-\infty}^x dx' \, J_0^C (x')\Big] \, = \, 0,
\end{eqnarray}
where $\alg{J}^A$ are Noether currents for the global (left or right) group multiplication.

The classical integrability of the Green-Schwarz superstring sigma
model in the \AdSxS \ background has been established in
\cite{Bena:2003wd}. The corresponding infinite set of
nonlocal classically-conserved charges is found according
to a logic very close to the one described above (similar
observations for the bosonic part of the action were made in
\cite{Mandal:2002fs}). Further work can be found
in
\cite{Alday:2003zb,Arutyunov:2003rg,Hatsuda:2004it,Das:2004hy,Alday:2005gi,Frolov:2005dj,Das:2005hp,Vicedo:2009sn}.

{\scriptsize We conclude with a remark on the Hopf
algebra structure of the nonlocal charges. How charges
(\ref{nnl}) can give rise to the coproduct (\ref{cop})
is shown in
\cite{Bernard:1992mu}. A
semiclassical treatment \cite{Luscher:1977rq,MacKay:1992rc} is as follows. One
imagines two well-separated solitonic excitations
as the classical version of a scattering state. Soliton $1$ is
localized in the region $(-\infty,0)$, soliton $2$
in $(0,\infty)$. Defining the {\it semiclassical action}
of a charge on such solution as evaluation on the
profile, one splits the current-integration in
individual domains relevant for each of the two
solitons, respectively:
\begin{eqnarray}
\alg{J}^A_{|profile} &=& \int_{-\infty}^{\infty} dx \, {J_0^A}_{|profile}
=\int_{-\infty}^{0} dx \, J_0^A \, + \, \int_{0}^{\infty} dx \, J_0^A \, \, \, \longrightarrow \, \, \,
\Delta (\alg{J}^A) = \alg{J}^A \otimes \mathbbmss{1} + \mathbbmss{1} \otimes \alg{J}^A,\nonumber\\
\widehat{\alg{J}}^A_{|profile} &=& \Big[ \int_{-\infty}^{0} dx \, J_1^A + \frac{1}{2} \, f^A_{BC} \, \int_{-\infty}^{0} dx \, J_0^B (x) \, \int_{-\infty}^{x}  dy \, J_0^C (y) \Big] \,  \nonumber\\
&& + \Big[ \, \int_{0}^{\infty} dx \, J_1^A  \, + \, \frac{1}{2} \, f^A_{BC} \, \int_{0}^{\infty} dx \, J_0^B (x) \, \int_{0}^{x} dy \,  J_0^C (y)\Big] \nonumber\\
&&+ \, \frac{1}{2} \, f^A_{BC} \, \int_{0}^{\infty} dx \, J_0^B (x) \, \int_{-\infty}^{0} dy \, J_0^C (y).
\end{eqnarray}
Upon quantization in absence of anomalies this gives (\ref{cop}) on the Hilbert space.\par
}

\subsection{The centrally-extended \texorpdfstring{$\alg{psu}(2|2)$}{psu(2|2)} Yangian}

In the previous section, we have described how algebraic structures related to integrability arise at the two perturbative ends of the AdS/CFT correspondence. To fully exploit these powerful symmetries one needs to take a further step, which allows to go beyond the perturbative regimes. One introduces the choice of a vacuum state, and considers excitations upon this vacuum. This choice breaks the full $\alg{psu}(2,2|4)$ symmetry down to a subalgebra. The excitations carry the quantum numbers of the unbroken symmetry, and they scatter {\it via} an integrable S-matrix. 

The choice that is normally made is, for instance, to consider a string (composite operator) of $Z$ fields (one of the three complex combinations of the six scalar fields of ${\cal{N}}=4$ SYM) as the vacuum state. The unbroken symmetry consists then of two copies of the $\alg{psu}(2|2)$ Lie superalgebra, which receive central extensions through quantum corrections. The same algebra appears on the string theory side. The excitations carrying the unbroken quantum numbers are called {\it magnons}, in analogy to the theory of spin-chains and magnetism. 

\subsubsection{The Hopf algebra of the S-matrix}
Upon choosing a vacuum, the residual symmetry carried by the magnon excitations is (two copies of) the centrally extended $\alg{psu}(2|2)$ Lie superalgebra (or $\alg{psu}(2|2)_c$):
\begin{eqnarray}
\begin{array}{ll}
\ [\mathbb{L}_{a}^{\ b},\mathbb{J}_{c}] = \delta_{c}^{b}\mathbb{J}_{a}-\frac{1}{2}\delta_{a}^{b}\mathbb{J}_{c}, &   \ [\mathbb{R}_{\alpha}^{\ \beta},\mathbb{J}_{\gamma}] = \delta_{\gamma}^{\beta}\mathbb{J}_{\alpha}-\frac{1}{2}\delta_{\alpha}^{\beta}\mathbb{J}_{\gamma},\\[1ex]
\ [\mathbb{L}_{a}^{\ b},\mathbb{J}^{c}] = -\delta_{a}^{c}\mathbb{J}^{b}+\frac{1}{2}\delta_{a}^{b}\mathbb{J}^{c}, & \ [\mathbb{R}_{\alpha}^{\ \beta},\mathbb{J}^{\gamma}] = -\delta^{\gamma}_{\alpha}\mathbb{J}^{\beta}+\frac{1}{2}\delta_{\alpha}^{\beta}\mathbb{J}^{\gamma},\\[1ex]
\ \{\mathbb{Q}_{\alpha}^{\ a},\mathbb{Q}_{\beta}^{\
b}\}=\epsilon_{\alpha\beta}\epsilon^{ab}\mathbb{C},&\ \{\mathbb{S}^{\ \alpha}_{a},\mathbb{S}^{\ \beta}_{b}\}=\epsilon^{\alpha\beta}\epsilon_{ab}\mathbb{C}^{\dag},\\[1ex]
\ \{\mathbb{Q}_{\alpha}^{a},\mathbb{S}^{\beta}_{b}\} =
\delta_{b}^{a}\mathbb{R}_{\alpha}^{\ \beta} +
\delta_{\alpha}^{\beta}\mathbb{L}_{b}^{\ a}
+\frac{1}{2}\delta_{b}^{a}\delta_{\alpha}^{\beta}\mathbb{H}.&
\end{array}
\end{eqnarray}
The generators $\mathbb{R}_{\alpha}^{\ \beta}$ and $\mathbb{L}_{a}^{\ b}$ form the two $\alg{su}(2)$ subalgebras which, together with the central elements $\{\mathbb{H},\mathbb{C},\mathbb{C}^\dag \}$, form the bosonic part of $\alg{psu}(2|2)_c$. The names are reminiscent of the unbroken $R$- and Lorentz symmetry of the model. The fermionic part is generated by the supercharges $\mathbb{Q}_{\alpha}^{a}$ and $\mathbb{S}^{\beta}_{b}$. The `dagger' symbol is to remember that, in unitary representations, the two charges are indeed conjugate of each other, and a similar conjugation condition holds for the supercharges.

The representation of \cite{Beisert:2005tm} gives a {\it dynamical} spin-chain, {\it i.e.} sites can be created/de\-stroyed by the action of the generators.
The central charges act as
\begin{eqnarray}
\mathbb{H} \, |p\rangle \, = \, \epsilon (p) \, |p\rangle, \, \, \, \, \,
\mathbb{C} \, |p\rangle \, = \, c (p) \, |p \, Z^-\rangle, \, \, \, \, \, \mathbb{C}^{\dag} \, |p\rangle \, = \, \bar{c} (p) \, |p \, Z^+\rangle,
\end{eqnarray}
where $Z^{+ (-)}$ adds (removes) one `site' ({\it i.e.}, one of the scalar fields $Z$ in the infinite string that constitutes the vacuum state) to (from) the chain. We denote as $|p\rangle$ the one-magnon state of momentum $p$. This state is given by $|p\rangle = \sum_n \, e^{i p n} \, |\cdots Z \, Z \, \phi(n) \, \, Z \cdots \rangle$, $\phi$ being one of the $4$ possible orientations of the `spin' in the fundamental representation of $\alg{psu}(2|2)_c$. The eigenvalue $\epsilon (p)$ is the energy (dispersion relation) of the magnon excitation. As we will shortly see, $c (p)$ contains the exponential of the momentum $p$ itself. So does $\bar{c} (p)$, which in unitary ({\it alias}, real-momentum) representations is just the conjugate of $c (p)$.

The length-changing property can be interpreted, at the Hopf
algebra level, as a nonlocal modification of the (otherwise
trivial) coproduct \cite{Gomez:2006va,Plefka:2006ze}. Let us spell out the case of the central charges.
When acting on a two-particle state, one computes
\begin{eqnarray}
&&\mathbb{C} \otimes \mathbbmss{1} \, |p_1\rangle \otimes |p_2\rangle = \nonumber\\
&&\mathbb{C} \otimes \mathbbmss{1} \, \sum_{n_1 << n_2} \, e^{i \, p_1 \, n_1 \, + \, i \, p_2 \, n_2} \, |\cdots Z \, Z \, \phi_1 \, \underbrace{Z \cdots Z}_{n_2 - n_1 - 1} \, \phi_2 \, Z \cdots \rangle  \, =\nonumber\\
&& \qquad \qquad \, \, \, ({\rm rescaling} \, \, n_2) \, \, = \, \, c (p_1) \, e^{i p_2}  \, |p_1\rangle \otimes |p_2\rangle.
\end{eqnarray}
This action is non-local, since
acting on the first magnon (with momentum $p_1$) produces a result
which also depends on the momentum $p_2$ of the second magnon. 

We must now impose compatibility of the S-matrix with the symmetry algebra carried by the excitations. 
Imposing such
S-matrix invariance condition $\Delta (\mathbb{C}) {\rm S} =
{\rm S} \Delta (\mathbb{C})$ implies computing
\begin{eqnarray}
\label{smatcop}
{\rm S} \, \Delta(\mathbb{C}) \, = {\rm S} \, [\mathbb{C} \otimes \mathbbmss{1} + \mathbbmss{1} \otimes \mathbb{C}] \, = {\rm S} \, [e^{i p_2} \mathbb{C}_{local} \otimes \mathbbmss{1} + \mathbbmss{1} \otimes \mathbb{C}_{local}],
\end{eqnarray}
where $\mathbb{C}_{local}$ is the {\it local} part of
$\mathbb{C}$, acting as $\mathbb{C}_{local} |p\rangle = c(p)
|p\rangle$. An analogous argument works for $\Delta (\mathbb{C}) {\rm S}$. One can rewrite (\ref{smatcop}) as
\begin{eqnarray}
\label{coprodo}
\Delta (\mathbb{C}_{local}) = \mathbb{C}_{local} \otimes e^{i p} + 1 \otimes \mathbb{C}_{local}.
\end{eqnarray}
Formula (\ref{coprodo}) is the manifestation of a
non-trivial Hopf-algebra coproduct\footnote{We remark that
a (nonlocal) basis change for spin-chain states can produce $e^{i p}$ factors in different places in the coproduct
(possibly with a different power), with no deep consequences.}. Similarly, to all (super)charges of $\alg{psu}(2|2)_c$, one
assigns an additive quantum number $[[A]]$ s.t.
\begin{eqnarray}
\label{coprodot}
\Delta (\alg{J}^A) = \alg{J}^A \otimes e^{i [[A]] p} + \mathbbmss{1} \otimes \alg{J}^A,
\end{eqnarray}
which gives a (Lie) superalgebra
homomorphism. Counit and antipode are derived from the Hopf algebra axioms, and the whole structure defines a consistent Hopf algebra. The S-matrix invariance should be written as
\begin{eqnarray}
\Delta^{op} R = R \Delta
\end{eqnarray}
(quasi-cocommutativity), where the invertible $R$-matrix is defined as $R=P {\rm S}$, $P$ being the graded permutation.
There is a consistency requirement: since $\Delta (\mathbb{C})$ is central,
\begin{eqnarray}
\label{coco}
\Delta^{op} (\mathbb{C}) \, R \, = \, R \, \Delta (\mathbb{C}) \, = \, \Delta (\mathbb{C}) \, R \qquad \implies \qquad
\Delta^{op} (\mathbb{C}) \, = \, \Delta (\mathbb{C}).
\end{eqnarray}
This is guaranteed by interpreting as algebraic condition the physical requirement
\begin{equation}
\label{interpr}
U \equiv e^{i p} \, \mathbbmss{1}= \kappa \, \mathbb{C} \, \, + \, \mathbbmss{1}
\end{equation}
for a constant $\kappa$ related to the coupling $g$
\cite{Beisert:2005tm}.

A version of the coproduct (\ref{coprodot}) was shown to emerge from the
dual worldsheet string-theory. In \cite{Klose:2006zd}, the result was
reproduced by applying the standard Bernard-LeClair procedure
\cite{Bernard:1992mu} to the light-cone worldsheet Noether
charges obtained in \cite{Arutyunov:2006ak}.

{\scriptsize A semi-classical
argument, based on the same reasoning presented
at the end of section 3, is as follows. The light-cone worldsheet
Noether supercharges have nonlocal contributions in the
physical fields:
\begin{eqnarray}
\alg{J}^A = \int_{-\infty}^{\infty} d \sigma \, J_0^A (\sigma ) \,
e^{i \, [[A]] \, \int_{-\infty}^{\sigma} \, d \sigma' \, \partial x^-
(\sigma')}.
\end{eqnarray}
If we consider, as before, two well-separated soliton excitations, the {\it semiclassical action} of these charges
on such a scattering state is again obtained by
splitting the integrals:
\begin{eqnarray}
{\alg{J}^A}_{|profile} &=&  \int_{-\infty}^{\infty} d \sigma \, J_0^A (\sigma )_{|profile}\,  e^{i \, [[A]] \, \int_{-\infty}^{\sigma} \, d \sigma' \, \partial x^- (\sigma')_{|profile}} \,  \nonumber\\
&=&\int_{-\infty}^{0} d \sigma \, J_0^A (\sigma ) e^{i \, [[A]] \, \int_{-\infty}^{\sigma} d \sigma' \, \partial x^- (\sigma')} \,  + \, \int_{0}^{\infty} d \sigma \, J_0^A (\sigma ) \, e^{i \, [[A]] \, \int_{-\infty}^{0} d \sigma' \, \partial x^- (\sigma')} \, e^{i \, [[A]] \, \int_{0}^{\sigma} d \sigma' \, \partial x^- (\sigma')}\,  \nonumber\\
&\sim &\alg{J}^A_1 \, + e^{i [[A]] \, p_1} \alg{J}^A_2 \,\, \, \, \,  \longrightarrow \, \,
\Delta (\alg{J}^A) = \alg{J}^A \otimes \mathbbmss{1} +  e^{i [[A]] \, p}\, \otimes \alg{J}^A, 
\end{eqnarray}
where one has used the definition of the worldsheet momentum
for the first excitation.\par }

From the Hopf-algebra antipode $\Sigma$ one derives
derive the so-called `antiparticle'
representation $\tilde{\alg{J}}^A$ and the corresponding
charge-conjugation matrix $C$:
\begin{eqnarray}
\label{pode} \Sigma (\alg{J}^A) \, = \, C^{-1} \, [ \,
\tilde{\alg{J}}^{A} ]^{st} \, C,
\end{eqnarray}
where $M^{st}$ is the supertranspose of $M$.
These are the ingredients entering the crossing-symmetry relations
originally written down in \cite{Janik:2006dc}, where the
existence of an underlying Hopf-algebra of the S-matrix was
conjectured. The antiparticle representation and the constraints on the overall scalar factor of the S-matrix as found in \cite{Janik:2006dc}, naturally follow from (\ref{pode}) combined with the general formulas
\begin{eqnarray}
(\Sigma \otimes \mathbbmss{1}) \, R \, = \, (\mathbbmss{1} \otimes \Sigma^{-1}) \, R \,= \, R^{- 1},
\end{eqnarray}
where the antipode is derived from the coproduct (\ref{coprodot}).

A reformulation in terms of a
Zamolodchikov-Faddeev (ZF) algebra has been given in
\cite{Arutyunov:2006yd}. There, the basic objects
are creation and annihilation operators, with
commutation relations given in terms of the S-matrix. Also, a $q$-deformation of this structure
and of the one-dimensional Hubbard model is studied in \cite{Beisert:2008tw,Beisert:2010kk}.

\subsubsection{The Yangian of the S-matrix}\label{sec:YS}
The S-matrix in the fundamental representation has been shown to
possess $\alg{psu}(2|2)_c$ Yangian symmetry \cite{Beisert:2007ds}.
In order to be a Lie superalgebra homomorphism, the coproduct should
respect (\ref{rels}). Therefore, the structure of the Yangian
coproduct has to take into account the deformation in
(\ref{coprodot}):
\begin{eqnarray}
\label{Niklas}
\Delta (\widehat{\alg{J}}^A) = \widehat{\alg{J}}^A \otimes \mathbbmss{1} + U^{[[A]]} \, \otimes \widehat{\alg{J}}^A + \frac{1}{2} \, f^A_{BC} \, \alg{J}^B \, U^{[[C]]} \, \otimes \alg{J}^C.
\end{eqnarray}
The representation for $\widehat{\alg{J}}^A$ is the so-called {\it evaluation} representation, typically obtained by multiplying level-zero generators by a `spectral' parameter. Here
\begin{eqnarray}
\label{u}
\widehat{\alg{J}}^A \, = \, u \, \alg{J}^A \, = \, i g \, (x^+ \, + \frac{1}{x^+} \, - \frac{i}{2 g}) \, \alg{J}^A.
\end{eqnarray}
The variables $x^\pm$ parameterize the fundamental
representation (conventions as in \cite{Arutyunov:2009mi}).

A special remark concerns the dual structure constants $f^A_{BC}$.
They should reproduce the general form (\ref{cop}), and analogous
ones with all indices lowered should be used to prove the Serre
relations (\ref{Serr}). However, since the Killing form of
$\alg{psu}(2|2)_c$ is zero, one has a problem in defining these
structure constants. In \cite{Beisert:2007ds}, the quantities
$f^A_{BC}$ are explicitly given as a list of numbers, without
necessarily referring to an index-lowering procedure\footnote{An
argument in \cite{Beisert:2007ds} suggests interpreting these
quantities as dual structure constants in an enlarged algebra with
invertible Killing form, see also
\cite{FabianThesis,Spill:2008zz}. This algebra is obtained by
adjoining the $\alg{sl}(2)$ automorphism of $\alg{psu}(2|2)_c$
\cite{Serganova,Beisert:2006qh}. Apart from allowing inversion of
the Killing form and determination of $f^A_{BC}$, these extra
generators would drop out of the final form of the Yangian
coproduct (\ref{Niklas}). We also refer to \cite{Matsumoto:2008ww} for a derivation of the Yangian coproducts using the exceptional Lie superalgebra ${\alg{D}}(2,1;\alpha)$.}. The table of coproducts is in this way
fully determined.

Another remark concerns the dependence of the spectral parameter
$u$ on the representation variables $x^\pm$, or, equivalently, on
the eigenvalues of the central charges of $\alg{psu}(2|2)_c$. For
simple Lie algebras, the spectral parameter is typically an
additional variable attached to the evaluation representation.
Together with the existence of a {\it shift}-automorphism $u
\rightarrow u + const$ of the Yangian in evaluation
representations, this implies that the Yangian-invariant S-matrix
is of difference-form ${\rm S} = {\rm S}(u_1 - u_2).$ The
dependence of $u$ on the central charges alters this property, and
one does not have a difference form in the fundamental S-matrix
(see \cite{BazhanovTalk} and section 3.1.1).

The full
quantum S-matrix is also invariant under the following exact symmetry, found in
\cite{Matsumoto:2007rh} and shortly afterwards confirmed in \cite{Beisert:2007ty}:
\begin{eqnarray}
\label{bb}
&&\Delta (\widehat{\mathbb{B}'})= \widehat{\mathbb{B}'} \otimes \mathbbmss{1} \, + \, \mathbbmss{1} \otimes \widehat{\mathbb{B}'}
+\frac{i}{2g} (\mathbb{S}^\alpha_{a} \otimes \mathbb{Q}^a_{\alpha} \, + \, \mathbb{Q}^a_{\alpha} \otimes \mathbb{S}^\alpha_{a} ),\nonumber\\
&&\Sigma (\widehat{\mathbb{B}'}) \, = \, - \widehat{\mathbb{B}'} \, + \, \frac{2i}{g} \mathbb{H},\nonumber\\
&&\widehat{\mathbb{B}'} \, = \, \frac{1}{4} (x^+ + x^- - 1/x^+ - 1/x^-) \, diag(1,1,-1,-1).
\end{eqnarray}
This coproduct is reminiscent of a level one
Yangian symmetry (cf. (\ref{cop})). We will see in the next section the relevance of this generator for the classical $r$-matrix. Commuting this symmetry with the (level zero) generators, one obtains novel exact Yangian (super)symmetries of S \cite{Matsumoto:2007rh}. The latter act on bosons and fermions with two {\it different} spectral parameters, reducing in the classical limit to the supercharges of \cite{Moriyama:2007jt}.

\section{The classical \texorpdfstring{$r$}{r}-matrix}
The form of the Yangian we discussed resembles the standard one while simultaneously showing some unexpected features. In order to gain a deeper understanding it is commonly advantageous to study certain limits. One important instance is the {\it classical} limit, {\it i.e.} one studies
perturbations of the $R$-matrix around the identity:
\begin{eqnarray}
\label{classexp}
R = \mathbbmss{1}\otimes \mathbbmss{1} \, + \,
\hbar \, r \, + \, {\cal{O}} (\hbar^2 ),
\end{eqnarray}
$\hbar$ being a small parameter. The first-order term $r$ is called the {\it classical} $r$-matrix\footnote{$r$ lives in $\alg{g}\otimes \alg{g}$, for $\alg{g}$ an algebra, $R$ in $U(\alg{g})\otimes U(\alg{g})$, $U(\alg{g})$ the universal enveloping algebra of $\alg{g}$.}. One can easily prove that, if $R$ satisfies the Yang-Baxter equation (YBE), $r$ satisfies the {\it classical} YBE (CYBE):
\begin{eqnarray}
\label{cYBE}
[r_{12},r_{13}] + [r_{12},r_{23}] + [r_{13},r_{23}] =0.
\end{eqnarray}
In known cases, studying (\ref{cYBE}) one can classify the solutions of the YBE itself, and the possible quantum group structures underlying such solutions (Belavin-Drinfeld theorem \cite{BD1,BD2}).
We will not reproduce here the details. Knowing the $r$-matrix, there is a standard procedure for constructing an associate Lie bialgebra, and {\it quantizing} it\footnote{Meaning completing the Lie bialgebra to a quantum group (classical $r$- to quantum R-matrix).} in terms of so-called `Manin triples' (see {\it e.g.} \cite{Etingof}). The quantum structures for simple Lie algebras are elliptic quantum groups, (trigonometric) quantum groups and Yangians. Analogous theorems for superalgebras are investigated in \cite{Leites:1984pt,Karaali1,Karaali2}. An illuminating example is Yang's $r$-matrix ($C_2$ is the quadratic Casimir) 
\begin{eqnarray}
\label{Yang}
r=\frac{C_2}{u_2 - u_1}~.
\end{eqnarray}
This is the prototypical rational solution of the CYBE\footnote{Since by definition $[C_2,\alg{J}^A \otimes \mathbbmss{1} + \mathbbmss{1} \otimes \alg{J}^A]=0 \, \, \forall A$, the CYBE is easily proven for (\ref{Yang}).}. The geometric series gives
\begin{eqnarray}
\label{Tayl}
&&r=\frac{C_2}{u_2 - u_1} = \frac{\alg{J}^A \otimes \alg{J}_A}{u_2 - u_1}=\sum_{n\geq 0} \alg{J}^A u_1^n \otimes \alg{J}_A u_2^{-n-1}=\sum_{n\geq 0} \alg{J}^A_n \otimes \alg{J}_{A,-n-1},
\end{eqnarray}
for $|u_1/u_2|<1$). Such rewriting attributes dependence on the $u_1$ ($u_2$) to operators in the first (second) space ({\it factorization}). This gives $r$ the form of tensor product of algebra representations. Assigning $\alg{J}^A_n \, = \, u^n \, \alg{J}^A$ in (\ref{Tayl}) gives loop-algebra relations
\begin{eqnarray}
\label{loopa}
[\alg{J}^A_m , \alg{J}^B_n] = \, f^{AB}_C \, \alg{J}^C_{m+n}.
\end{eqnarray}
The loop algebra is precisely the `classical' limit of the Yangian
$\Ya{\alg{g}}$ (see section \ref{ssec:drinf1}). With this example
one realizes how {\it rational} solutions of the CYBE, such as
(\ref{Yang}), starting as not-better specified elements of
$\alg{g}\otimes \alg{g}$ for a Lie algebra $\alg{g}$, give rise to
Yangians upon quantization (namely, their quantized version takes
values in $\Ya{\alg{g}}\otimes \Ya{\alg{g}}$). For related aspects
concerning the classical $r$-matrix, see
\cite{Magro:2010jx,SchaferNameki:2010jy}.

\subsection{\texorpdfstring{$\alg{psu}(2|2)_c$}{psu(2|2)c}}\label{ssec:clpsu}
In the case of the S-matrix found in \cite{Beisert:2005tm}, the parameter controlling the classical expansion is naturally the inverse of the coupling constant $g$ (near-BMN limit \cite{Berenstein:2002jq}):
\begin{eqnarray}
R = \mathbbmss{1} \otimes \mathbbmss{1} \, + \, \frac{1}{g} \, r
\, + {\cal{O}}(\frac{1}{g^2}).
\end{eqnarray}
The classical $r$-matrix $r$ is identified with the tree-level string scattering matrix computed in \cite{Klose:2006zd}. In the parameterization of  \cite{Arutyunov:2006iu} one has
\begin{equation}
\label{clAF}
x^\pm (x) = x \sqrt{1 - \frac{1}{g^2 (x - \frac{1}{x})^2}} \pm \frac{i x}{g (x - \frac{1}{x})} \, \, \to \, \, x.
\end{equation}
One sends $g$ to $\infty$ with $x$
fixed. $x$ is interpreted as an unconstrained
`classical' variable. This classical limit was studied in
\cite{Torrielli:2007mc}. The target is finding
the complete algebra the $r$-matrix takes values in, whose
quantization can reveal the full quantum symmetry of the
S-matrix. The fundamental representation tends to a limiting centrally-extended $\alg{psu}(2|2)$,
with generators parameterized by $x$. The classical $r$-matrix
$r=r(x_1,x_2)$ is not of difference form. The Lie
superalgebra is not simple and has zero dual Coxeter number. This prevents applying Belavin-Drinfeld type of theorems. Nevertheless,
$r$ has a simple pole at $x_1 - x_2 =
0$ with residue\footnote{As a consequence of the CYBE, such residue must be a Casimir.} the Casimir $C_2$ of
$\alg{gl}(2|2)$:
\begin{eqnarray}
C_2 = \sum_{i,j=1}^4 \, (-)^{[j]} \, E_{ij}\otimes E_{ji},
\end{eqnarray}
with $E_{ij}$ matrices with all zeros but $1$ in
position $(i,j)$, and $[j]$ the fermionic grading of the index
$j$. In the absence of a quadratic
Casimir for $\alg{psu}(2|2)_c$, the classical $r$-matrix displays
on the pole (it
`borrows') the Casimir of a bigger
algebra\footnote{$\alg{gl}(2|2)$ is obtained by adjoining to
$\alg{su}(2|2)$ the non-supertraceless element
$\mathbb{B}=diag(1,1,-1,-1)$.} for which a non-degenerate form exists and
the quadratic Casimir can be
constructed. This `borrowing' reminds a
mathematical prescription due to Khoroshkin and
Tolstoy \cite{KT,Khoroshkin:1994uk}. One expects that, if a universal
$R$-matrix exists and if it has
to be of Khoroshkin-Tolstoy type, an additional
Cartan element of type $\mathbb{B}$ has to appear.

Type-$\mathbb{B}$ generators
play an important role in factorizing $r$. The present $r$ is more complicated than Yang's one, and it is harder to find a
suitable geometric-like series expansion. A
first proposal for the fundamental representation was given \cite{Moriyama:2007jt},
with a Yangian tower of $\mathbb{B}$'s coupled to a tower of $\mathbb{H}$'s
to achieve factorization. This proposal fails to reproduce the
bound-state classical $r$-matrix \cite{MariusThesis}.

A universal formula was advanced in \cite{Beisert:2007ty}. It has been shown to reproduce also the classical limit of the bound-state S-matrix \cite{deLeeuw:2008dp,Arutyunov:2009mi}, and it reads
\begin{eqnarray}
\label{eqn;Rmat}
r = \frac{\mathcal{T}-\tilde{\mathbb{B}}\otimes
\mathbb{H}-\mathbb{H}\otimes
\tilde{\mathbb{B}}}{i(u_{1}-u_{2})}-\frac{\tilde{\mathbb{B}} \otimes \mathbb{H} }{iu_{2}}
+\frac{\mathbb{H}\otimes \tilde{\mathbb{B}}}{iu_{1}} - \frac{\mathbb{H}\otimes \mathbb{H}}{\frac{2 i u_1 u_2}{u_1 - u_2}},
\end{eqnarray}
\begin{eqnarray}
&&\mathcal{T}=2\left(\mathbb{R}^{\
\alpha}_{\beta}\otimes\mathbb{R}^{\ \beta}_{\alpha}- \mathbb{L}^{\
a}_{b}\otimes\mathbb{L}^{\ b}_{a}+ \mathbb{S}^{\
\alpha}_{a}\otimes\mathbb{Q}^{\ a}_{\alpha}- \mathbb{Q}^{\
a}_{\alpha}\otimes\mathbb{S}^{\ \alpha}_{a}\right),\nonumber\\
&&\tilde{\mathbb{B}} =\frac{1}{4 \, \epsilon (p)} \, \, diag (1,1,-1,-1).
\end{eqnarray}
In this formula, the generators are in their classical limit, the variable $u$ is the classical limit of (\ref{u}), and $\epsilon (p)$ is the classical energy (cf. section \ref{ssec:clpsu}). All classical Yangian generators are obtained as $\alg{J}_n = u^n \alg{J}$ after factorizing {\it via} the geometric series expansion. Quantization of this formula is an open problem. The classical analysis seems to suggest that the triple central extension may have to merge into some sort of deformation of the loop algebra of $\alg{gl}(2|2)$, where the additional generator $\mathbb{B}$ is sitting. Another open question is how to relate the results described here to the $r,s$ non-ultralocal structure of the $\alg{psu}(2,2|4)$ sigma-model \cite{Magro:2010jx,SchaferNameki:2010jy,Vicedo:2010qd}.

\subsubsection{Difference Form}
Formula (\ref{eqn;Rmat}) displays an interesting structure where the dependence on the
spectral parameter $u$ is (almost purely) of difference form. The non-difference form is encoded in the representation labels
$x^\pm (u)$ appearing in the symmetry generators, and in the last three terms of formula (\ref{eqn;Rmat}).
Moreover, Drinfeld's second realization for the $\alg{psu}(2|2)_c$ Yangian has been obtained in \cite{Spill:2008tp},
together with the suitable evaluation representation. The Yangian Serre relations, which were left as an open question in \cite{Beisert:2007ds},
are proven to be satisfied in the second realization (see also \cite{Matsumoto:2009rf}.) The representation of \cite{Spill:2008tp}
possesses a shift-automorphism $u \rightarrow u + const$,
which normally guarantees the difference form of the S-matrix.
All this suggests the following, provided an algebraic interpretation of the last three terms in formula (\ref{eqn;Rmat}) can be found that generalizes to the full quantum case (possibly along the case of the ideas reported in \cite{Beisert:2007ty} in terms of twists). Modulo this interpretation, one might hope to achieve a rewriting of
the quantum S-matrix such that the dependence on $u_1$ and $u_2$
is (almost purely) of difference form, the rest being taken care of
by suitable combinations of algebra generators\footnote{In the
fundamental representation, such a rewriting has been shown to be
possible in \cite{Torrielli:2008wi}. The resulting form is
reminiscent of what a Khoroshkin-Tolstoy type of formula (or some natural quantization of the classical
$r$-matrix (\ref{eqn;Rmat})) would look like in this
representation.}. One would expect this as the result of
evaluating a hypothetical Yangian universal $R$-matrix in this
particular representation. This expectation seems to be consistent
with recent studies of the
exceptional Lie superalgebra ${\alg{D}}(2,1;\alpha)$
\cite{Beisert:2005tm,Matsumoto:2008ww,Matsumoto:2009rf}\footnote{$\alg{psu}(2|2)_c$
can be obtained by suitable contraction of ${\alg{D}}(2,1;\alpha)$. See also \cite{Heckenberger:2007ry}.}, and with the explicit form of the bound state S-matrix (see next section).

\section{The bound state S-matrix}
The previous discussion highlights the importance of investigating the structure of the S-matrix for generic representations of $\alg{psu}(2|2)_c$. One motivation is obtaining the universal $R$-matrix and understanding the role of the $\widehat{\mathbb{B}}'$ symmetry. There is also a more stringent need related to finite-size corrections to the energies according to the TBA approach \cite{Bajnok:2010ke}. According to this philosophy, it becomes crucial to have a concrete realization of the (mirror) bound state S-matrices. Usually, these can be {\it bootstrapped} once the S-matrix of fundamental constituents is known \cite{Zamolodchikov:1978xm,Dorey:1996gd}. However, the present case is more complicated. The fundamental S-matrix does not reduce to a projector on the bound state pole, related to the fact that the tensor product of two short representations (generically irreducible) becomes reducible but indecomposable on the pole. The only way to construct the S-matrix for bound states seems to be a direct derivation from the Lie superalgebra invariance in each bound state representation. This becomes rapidly cumbersome \cite{Arutyunov:2008zt}. Moreover, this does not uniquely fix the S-matrix when the bound state number increases, and one needs to resort to YBE, or, as shown in \cite{deLeeuw:2008dp}, to Yangian invariance. The Yangian eventually provides an efficient solution to this problem and it allows to uniquely determine the S-matrix for arbitrary bound state numbers \cite{Arutyunov:2009mi}.

The bound state representations are atypical (short) completely symmetric representations of dimension $4 \ell$, $\ell=1,2,...$. They all extend to evaluation representations of the Yangian, with appropriate evaluation parameter $u$ \cite{deLeeuw:2008dp}. A convenient realization is given in terms of differential operators acting on the space of degree $M$ polynomials (superfields) in two bosonic ($w_a, \, a=1,2$) and two fermionic
($\theta_\alpha, \, \alpha=1,2$) variables. All details can be found in \cite{Arutyunov:2009mi}. The essence of the construction consists in finding a closed subset of states $|x_i\rangle$ for which the S-matrix can be computed exactly in terms of a definite matrix $M$. One then generates all other states $|y_A\rangle$ by acting with (Yangian) coproducts on this closed subsector, and using quasi-cocommutativity:
\begin{eqnarray}
R |y_A\rangle \, = \, R \, \Delta (\mathbb{J})^i_A \, |x_i\rangle \, =  \, \Delta^{op} (\mathbb{J})^i_A \, R \, |x_i\rangle \, = \, \Delta^{op}(\mathbb{J})^i_A \, M^j_i \, |x_j\rangle.
\end{eqnarray}
On the other hand, $R |y_A\rangle = R^B_A \, |y_A\rangle = R^B_A \,\Delta (\mathbb{J})^i_B \, |x_i\rangle$. The task is to find as many states as needed to invert the above relation, namely $R^B_A =  \Delta^{op} (\mathbb{J})^i_A \, M^j_i \, [\Delta (\mathbb{J})^{-1}]^B_j$.

The construction automatically provides a {\it factorizing twist} \cite{twi}
for the R-matrix in the bound state representations (hence also for the fundamental representation):
\begin{eqnarray}
\label{factor}
R = F_{21} \times {F_{12}}^{-1}.
\end{eqnarray}
However, we remark that the coproduct twisted with $F_{12}$ is by construction cocommutative, but, as expected, not at all trivial.
Furthermore, apart perhaps from the overall factor, the bound state S-matrix depends only on $u_1 - u_2$, on combinatorial factors involving
the integer bound-state components, and on specific combination of
algebra labels $a_i,b_i,c_i,d_i$. These combinations are the same noticed in
\cite{Torrielli:2008wi}. It remains hard to figure out a
universal formula reproducing this S-matrix. Nevertheless, it looks like such a universal
object would treat the evaluation parameters of the Yangian as
truly independent variables, the latter appearing only in difference-form due
to the Yangian shift-automorphism. The rest of the labels would
appear because of the presence in the universal R-matrix of the (super)charges in the typical `positive
$\otimes$ negative'-roots combinations, breaking
the difference-form due to the constraint that links the
evaluation parameter to the central charges. This is consistent with the findings of \cite{Arutyunov:2009ce}, where one of the blocks of the S-matrix has been related to the universal R-matrix of the Yangian of $\alg{sl}(2)$ in arbitrary bound state representations.

The bound state S-matrix have been utilized in \cite{Arutyunov:2009iq} to verify certain conjectures appeared
in the literature, concerning the eigenvalues of the transfer matrix in specific short representations \cite{Beisert:2006qh}. Long representations have been studied in \cite{Arutyunov:2009pw}.

\section{Acknowledgments}
I would like to warmly thank C.\ Sieg, J.\ Plefka, F.\ Spill, H.\
Yamane, I.\ Heckenberger, S.\ Moriyama, T.\ Matsumoto, G.\ Arutyunov,
M.\ de Leeuw and R.\ Suzuki for the most enjoyable collaborations on
the topic of this review, and all the people I had the privilege
to share discussions with. I would like to especially thank N.\
Beisert for carefully reading the manuscript and suggesting
several improvements. This work has been partially
supported by EPSRC through the grant no. EP/H000054/1.

\phantomsection
\addcontentsline{toc}{section}{\refname}
\bibliography{chapters,intads,AlesYangian}

\begin{thebibliography}{10}
\bibitem{Vicedo:2010qd}
\nbbstauthor{B.~Vicedo},
\nbbsttitle{``The classical R-matrix of AdS/CFT and its Lie dialgebra
  structure''},
\nbbsteprint{\arxivref{1003.1192}{arxiv:1003.1192}}.
\end{thebibliography}
\bibliographystyle{nb}

\end{document}